\documentclass[superscriptaddress,twocolumn,prl,amsmath,amssymb]{revtex4}
\usepackage[varg]{txfonts}

\usepackage{graphicx}

\begin{document}

\title{Vortices and quasiparticles near the
``superconductor-insulator" transition in thin films}

\author{Victor~M.~Galitski} \affiliation{Kavli Institute for
Theoretical Physics University of California Santa Barbara, CA
93106-4030} \author{G.~Refael} \affiliation{Kavli Institute for
Theoretical Physics University of California Santa Barbara, CA
93106-4030} \author{Matthew~P.~A. Fisher} \affiliation{Kavli Institute
for Theoretical Physics University of California Santa Barbara, CA
93106-4030} \author{T.~Senthil} \affiliation{Center for Condensed
Matter Theory, Indian Institute of Science, Bangalore-560012, India}
\affiliation{Department of Physics, Massachusetts Institute of
Technology, Cambridge, Massachusetts 02139}

\begin{abstract}
We study the low temperature behavior of an amorphous
superconducting film driven normal by a perpendicular magnetic field
(B). For this purpose we introduce a new two-fluid formulation consisting of
fermionized field induced vortices and electrically neutralized
Bogoliubov quasiparticles (spinons) interacting via a long-ranged
statistical interaction.  This approach allows us to access a novel
non-Fermi liquid phase which naturally interpolates between the low B
superconductor and the high B normal metal.  We discuss the transport,
thermodynamic, and tunneling properties of the resulting ``vortex
metal" phase.
\end{abstract}

\maketitle
Superconductivity in two dimensions (2d) provides a unique area to
explore a fascinating variety of quantum phenomena. Of particular
interest are strongly disordered superconducting films, tuned by a
perpendicular magnetic field into the normal state.  Early
experiments~\cite{Goldman1, Hebard,Gantmakher} focused on a
magnetic-field tuned ``superconductor-insulator" transition at a
critical field $B_{\rm c}$.  Based on a phenomenological ``dirty boson"
model~\cite{Fisher}, universal scaling behavior in the temperature and
field dependence of the resistance was predicted near the quantum
phase transition ($B \approx B_{\rm c}$) separating the superconductor from
the Bose insulator.  Although some experimental support was found for
scaling of $R(T,B)$, the resistance at the separatrix, $R(T=0,B_{\rm c})$
was often found to deviate significantly from the expected universal
value  (near $R_Q = h/4e^2$).

More recent experiments on such amorphous films at lower temperatures
revealed a rather rich and surprising behavior in $R(T\rightarrow
0,B)$.  Specifically, for temperatures well below the mean field
transition where Cooper pairs are well formed, the resistance
saturates at a value which can be many orders of magnitude smaller
than the normal state resistance~\cite{Kapitulnik1,Kapitulnik2}.  This
mysterious {\em strange metal} phase occurs over a reasonably large
range of intermediate fields, $B \le B_{\rm c}$.  Moreover, in highly
disordered films with $B_{\rm c} \ll H_{\rm c2}$, the resistance at higher fields
was found to exhibit a dramatic non-monotonic dependence, rising in
some cases up to values of order $10^6 R_Q$, before plummeting towards
the normal state resistance at $B \approx H_{\rm c2}$~\cite{HPR, Shahar,KapitulnikSteiner}. A boson-only theory that might exhibit a metallic phase was proposed by Dalidovic et al. \cite{Phillips,PhillipsPRL}.  Kapitulnik et al.~\cite{Kivelson}, however, argued for the importance of fermionic quasiparticles ignored in the dirty-boson approach.

In this Letter we develop a two-fluid formulation, which incorporates
fermionic quasiparticles into the dirty-boson model.  The two fluids
consist of field induced vortices and electrically neutralized
quasiparticles (``spinons") interacting via a long-ranged statistical
interaction.  In the low temperature limit the vortices
must be treated as quantum particles and their statistics will be
important.  Previous work has implicitly assumed that the vortices are
bosonic.  However in the presence of unpaired electronic
excitations the issue of vortex statistics can be subtle.
In this paper we explore the consequences of treating the
vortices as fermions (see also \onlinecite{Larkin}).  Technically this is achieved via a Chern-Simons flux attachment to the bosonic vortices.
With fermionic vortices and fermionic electron-like quasiparticles our
two-fluid model can be studied within an RPA approximation.
The fermionized vortices can quantum diffuse, leading to a
description of a novel conducting non-Fermi liquid phase which
interpolates between the low field superconductor and the high field
normal state.  We suggest that this ``vortex metal" phase might
account for the {\it strange metallic} behavior observed in InO and
MoGe films.  Moreover, our two-fluid approach gives a natural
explanation for the non-monotonic dependence of the low temperature
magnetoresistance.  Below we discuss transport, thermodynamic, and
tunneling properties of the ``vortex metal.''

Consider a 2d disordered superconductor in a perpendicular magnetic
field below the mean-field BCS upper critical field $H_{\rm c2}$. For
concreteness we consider a lattice tight-binding  Hamiltonian of an
$s$-wave superconductor,
\begin{eqnarray}
\label{H}
\nonumber {\cal H} = \!\!\!&-& \!\!\! t \sum\limits_{\left\langle {\bf r} {\bf r}'
\right\rangle} \left[ c_{{\bf r} \alpha}^\dagger c_{{\bf r}' \alpha}
e^{i {\bf A}_{{\bf r} {\bf r}'}} + \mbox{h.~c.} \right] + \frac{1}{2C}
\sum\limits_{\bf r} \left( 2n_{\bf r} + c^\dagger_{{\bf r}\alpha}
c_{{\bf r}\alpha} - n_0 \right)^2 \\ \!\!\! &+& \!\!\! \Delta \sum\limits_{\bf r}
\left[ e^{i\phi_{\bf r}}  c_{{\bf r} \uparrow} c_{{\bf r} \downarrow}
+ \mbox{h.~c.} \right] + {\cal H}_{\rm disorder},
\end{eqnarray}
where  $\Delta$ is the magnitude of the superconducting order
parameter, $C^{-1}$ is an on-site charging energy and $n_0$ is the
electrical charge density.  At temperatures well below the mean-field
transition, it is necessary to include quantum fluctuations of the
Cooper pairs, which were incorporated phenomenologically via the phase
of the superconducting order parameter, $\phi_{\bf r}$, which is
conjugate to the Cooper pair number operator, $n_{\bf r}$.  Due to the
external magnetic field and strong disorder, one expects a substantial
number of low energy electrons will also be present, certainly in the
vortex cores and perhaps elsewhere.  Moreover, since the {\it strange
metal} is resistive, the vortices are mobile, and a correct
description will likely require incorporating the quantum dynamics of
{\it both} vortices and fermionic quasiparticles.  Such a formulation
has recently been obtained in Refs.~\cite{SenthilFisher,
BalentsFisher}, and will be reviewed and exploited herein.

Following Ref.~\cite{SenthilFisher} it is convenient to make a change
of variables: $f_{{\bf r} \alpha }^{\dagger} = b_{\bf r} c_{{\bf r}
\alpha }^{\dagger}$, where $\left( b^\dagger_{\bf r} \right)^2 = e^{i
\phi_{\bf r}}$ creates a Cooper pair at site ${\bf r}$. This
transformation removes the electric charge from the quasiparticles,
leaving charge zero and spin one-half fermionic ``spinons", $f_{\bf
r}$. The use of spinon variables is merely a technical convenience and
does not necessarily imply that they are the good excitations in the
system.  Indeed in the vortex metal phase the spin-carrying
excitations will ultimately be electrons (not spinons).  Next, a
duality transformation can be implemented which exchanges the
operators $b_{\bf r}$ for $hc/2e$ vortices, leading to a theory of
bosonic vortices in a fluctuating gauge field which mediates the
inter-vortex long-range interactions.  Moreover, the vortices and
spinons have a statistical interaction between them, {\em i.~e.}, a
vortex ``sees" a spinon as a source of a $\pi$-flux and vice versa.

We describe the vortex-spinon mixture with a so-called
$U(1)$ formulation~\cite{BalentsFisher,Read}.
In the long-wavelength low-energy limit the corresponding Euclidean
action is: $S = \int d{\bf x} d\tau [{\cal L}_{\rm v} +
{\cal L}_{\rm s} + {\cal L}_{\rm int}]$, where ${\cal L}_{\rm v}$ is
the vortex Lagrangian in terms of the vortex bosonic fields $\Psi$ and
the gauge field $a_\mu =( {\bf a},a_0)=(a_x,a_y,a_0)$,
\begin{equation}
\label{Sbv}
{\cal L}_{\rm v} = \Psi^\dagger \left[- \frac{\left( {\bm \nabla} -i
{\bf a} + i \bm{\alpha} \right)^2}{2 m_{\rm v}} +  \left(\partial_\tau - i
a_0 + i \alpha_0 \right) \right] \Psi + {\cal L}_a  ,
\vspace*{-0.05in}
\end{equation}
with: ${\cal L}_a = \frac{1}{2C} \left( \epsilon_{\mu \nu \lambda}
\partial_\nu a_\lambda  - \delta_{\mu \tau} \pi n_0 \right)^2$.  The
spinon's action is
\begin{equation}
\label{Ssp}
{\cal L}_{\rm s} =  f^\dagger_\alpha \left[- {1 \over 2 m_{\rm s}}
\left({\bm \nabla} - i {\bm \beta} \right)^2 +  \left(\partial_\tau -
i \beta_0 \right) \right] f_\alpha,
\vspace*{-0.05in}
\end{equation}
and the vortex-spinon statistical interaction is mediated by two
auxiliary $U(1)$ fields $\alpha_\mu$ and $\beta_\mu$,
\begin{equation}
{\cal L}_{\rm int} = -(i/\pi) \epsilon_{\mu \nu \lambda} \alpha_\mu
\partial_\nu \beta_\lambda.
\vspace*{-0.05in}
\end{equation}
The equations of motion $\delta {\cal L}/\delta \alpha_0=0$ and
$\delta {\cal L}/\delta \beta_0=0$ attach $\pi$ flux to the vortices
and spinons,
\begin{equation}
\epsilon_{ij} \partial_i \beta_j = \pi \hat{N}_{\rm v} = \pi \Psi^\dagger
\Psi ; \hskip1cm \epsilon_{ij} \partial_i \alpha_j = \pi \hat{n}_{\rm f} =
\pi  f^\dagger_\alpha f_\alpha .
\vspace*{-0.05in}
\end{equation}
with $\hat{N}_{\rm v}$ and $\hat{n}_{\rm f}$ the vortex and spinon densities.  The
total electrical charge density is given by $\hat{n}_0 =
(\epsilon_{ij} \partial_i a_j)/\pi$, and for small capacitance will be
set by the c-number $n_0$, that is $\langle \hat{n}_0 \rangle =
n_0$. The average number of vortices is set by the external magnetic
field $B$ through $\pi \left\langle N_{\rm v} \right\rangle = B$ (in units
where $\hbar = c = e = 1$).


In the {\it strange metal}, where the film resistance $R(T)$ saturates
at low temperatures, it appears that the vortices are diffusing.  To
access this within the vortex-spinon theory, it will prove extremely
convenient to statistically transmute the vortices, converting them
into fermions.  This can be achieved by attaching $2 \pi$
``statistical"  flux to each vortex, introducing a Chern-Simons gauge
field $A_\mu$.
If we denote the Lagrangian in Eq.~(\ref{Sbv}) as ${\cal
L}_{\rm v}(\Psi,a_\mu,\alpha_\mu)$, the Lagrangian for the fermionized
vortices (denoted as $\psi$) is simply,
\begin{equation}
{\cal L}_{{\rm {\rm fv}}} = {\cal L}_{\rm v}(\psi,a-A,\alpha) + {i \over 4 \pi} A_\mu
\epsilon_{\mu \nu \lambda} \partial_\nu A_\lambda .
\label{csaction}
\vspace*{-0.05in}
\end{equation}
As recently argued, fermionization of vortices is expected to be an
extremely good approximation due to the long-ranged intervortex
interaction~\cite{AliceaMotrunich}.  Indeed, by defining $a'_\mu =
a_\mu - A_\mu$, and then integrating over $A_\mu$, we absorb the
Chern-Simons gauge field $A_\mu$ into $a_\mu$.  One thereby obtains,
\begin{equation}
{\cal L}_{{\rm fv}} = {\cal L}_{\rm v}(\psi,a',\alpha)  + O(\partial^3 a'^2),
\vspace*{-0.05in}
\end{equation}
where we have dropped terms less important than the Maxwell term
present in ${\cal L}_{\rm v}$.  Remarkably, the resulting theory of
fermionized vortices has {\it no Chern-Simons term}.  The full
Lagrangian, ${\cal L}_{{\rm fv}} + {\cal L}_{\rm s} + {\cal L}_{\rm int}$, describes
fermionic vortices and spinons interacting via a statistical
interaction, and constitutes our two-fluid formulation of 2d
superconductors in a field.


Generally, the fermionized vortices will see an effective average
``dual'' magnetic field with strength
\begin{equation}
b_{{\rm fv}} = \pi \langle \hat{n}_0 - \hat{n}_{\rm f} - 2 \hat{N}_{\rm v}  \rangle .
\label{vortexb}
\vspace*{-0.05in}
\end{equation}
The electrical charge density and the vortex density are externally
determined conserved quantities. But now consider $\hat{n}_{\rm f}$: as it
stands, in addition to the conservation of total electrical charge,
the total spinon (or electron) number, $N_{\rm f} = \int d{\bf x} \langle
\hat{n}_{\rm f} \rangle$ is also conserved by ${\cal L}$.  This additional
global symmetry is present because we have dropped a term proportional
to $\Delta$,
\begin{equation}
{\cal L}_\Delta = \Delta [ f_{{\bf r} \uparrow} f_{{\bf r}\downarrow}
{\cal O}^\dagger_\alpha({\bf r}) + h.c.] ,
\label{Ldel}
\vspace*{-0.05in}
\end{equation}
where the operator $ {\cal O}^\dagger_\alpha({\bf r})$ inserts $2 \pi$
flux in $\epsilon_{ij} \partial_i \alpha_j$ at ${\bf r}$.  This term
originates microscopically from the term $e^{-i\phi}cc$ whereby a
Cooper pair is created from two electrons. As we will later show, the
term ${\cal L}_\Delta$ is perturbatively irrelevant in the ``vortex
metal" phase. Thus the conservation of $N_{\rm f}$ is an ``emergent
symmetry" not present microscopically.  Although the term in
Eq.~(\ref{Ldel}) is irrelevant, it has an important role: it allows
the spinon density to adjust, such that the total energy of the system
with $\Delta = 0$ is minimized.  This condition then determines the
spinon density $\hat{n}_{\rm f}$.  It is possible that over some range of
parameters the spinon  density adjusts itself to make $b_{{\rm fv}}=0$ and
fermionic vortices are natural variables. Another example is the limit
of vanishing vortex mass, $m_{\rm v} \rightarrow 0$. In this case, the
bosonic vortices ``condense" and expel the flux ($b_{\rm v}=0$), which sets
$\langle \hat{n}_{\rm f} \rangle = n_0$, effectively ``gluing" the charge
back on to the spinons. Thus, one recovers an ordinary Fermi liquid of
electrons. Alternately this may be described as an integer quantum
Hall state for fermionized vortices with $\nu_{\rm v}=1$.






Despite the {\em average} effective field in Eq.~(\ref{vortexb}), the
transverse force on a moving vortex is expected to be small in the
disordered situations of interest - essentially due to the ``normal"
core, the {\em local} dual magnetic field in the vicinity of the
vortex is small.  In this case, we expect that the fermionized
vortices can quantum diffuse, leading to a non-zero resistance at low
temperatures.
Below we explore the properties of the ``vortex metal" phase resulting
from the action in Eqs. (\ref{Sbv}-\ref{csaction}).  We neglect
possible quantum interference effects.


\begin{figure}[htbp]
\centering \includegraphics[width=6.2cm]{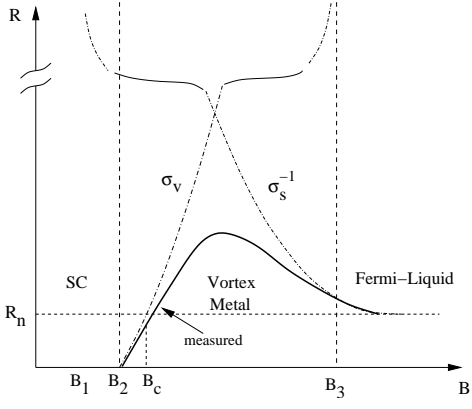}
\caption{Schematic $T=0$ phase diagram of amorphous superconducting
films in the two-fluid model, as a function of an applied magnetic
field.  Dashed curves label the vortex conductivity and spinon
resistivity, which determine via Eq.~(\ref{res}) the electrical
resistance (solid line).  The ``vortex metal" phase separates the low
field superconductor ($B<B_2$) from the high field Fermi liquid
($B>B_3$).  }
\label{fig1}
\end{figure}

For transport properties, a semi-classical Boltzmann equation or even
Drude theory is probably adequate.  In the latter case, the statistics
of the particles are not important, but it is essential that the
statistical interaction between vortices and spinons is correctly
incorporated.  Specifically, when the spinons move they induce an
electro-motive force (EMF) on the vortices and vice versa. Similarly,
an electrical charge current induces an EMF on the vortices. These
effects can be summarized in the following transport equations, which
describe the response of the system to an electrical current $\vec{\bf
J}$ and to a thermal gradient ${\bm \nabla} T$:
\begin{equation}
\label{qc}
\begin{array}{c}
\vec{\bf j}_{\rm v}=\sigma_{\rm v} \cdot  \hat{z}\times ( \vec{\bf
j}_{\rm s} -  \vec{\bf J} )-\lambda_{\rm v} {\bm \nabla} T;\\ \vec{\bf
j}_{\rm s}=\sigma_{\rm s} \cdot \hat{z}\times \vec{\bf j}_{\rm
v}-\lambda_{\rm s} {\bm \nabla} T,
\end{array}
\vspace*{-0.1in}
\end{equation}
where $\vec{\bf j}_{\rm v/s}$ are vortex and spinon currents,
respectively, and $\hat{z}$ is a unit vector normal to the film.
Here, $\sigma_{\rm v/s}$ and $\lambda_{\rm v/s}$ are dimensionless
transport ``conductivities" of the bosonic vortices/spinons
determining their current response to the EMF and thermal gradient.
Generally these quantities are two-by-two matrices with off-diagonal
components.  The resistivity tensor of the fermionic vortices
$(\sigma_{{\rm fv}})^{-1}$ will differ from those of the bosonic ones by
integer off-diagonal terms due to the Chern-Simons  flux attachment.
Thus, $(\sigma_{{\rm fv}})^{-1} = (\sigma_{\rm v})^{-1} + 2\pi\epsilon$, with
$\epsilon$ the unit antisymmetric tensor ($\epsilon_{xy} = 1$).  The
physical electrochemical potential gradient is given by ${\cal
\vec{E}} = \hat{z}\times \vec{\bf j}_{\rm v}$. Note that both the magnetic
field and the electric field seen by the spinons is the same as the
external ones. Thus the spinons respond to the external
electromagnetic field as electrons.  In the absence of a thermal
gradient one can extract the (dimensionless) electrical sheet
conductance matrix defined by, $J =\sigma {\cal E}$:
\begin{equation}
\sigma =   (\sigma_{\rm v})^{-1} + \sigma_{\rm s} .
\label{res}
\vspace*{-0.05in}
\end{equation}
Thus the spinon and vortex contributions to the total conductivity
simply add within this two-fluid theory.  It is expected that the
off-diagonal components of $\sigma_{\rm v}^{-1}$ and $\sigma_{\rm s}$ are both
small compared to the diagonal component. Ignoring them we obtain for
the longitudinal resistance $R = \sigma_{\rm v}^{xx}/(1+\sigma_{\rm v}^{xx}
\sigma_{\rm s}^{xx})$.

Before exploring the ``vortex metal" phase, we discuss the expected
($T=0$) behavior of $\sigma_{\rm v/s}$ as the strength of the external
magnetic field is varied (Fig.~\ref{fig1}).  At low field the vortices
are well separated and localized by impurities and the strong
inter-vortex interaction, implying $\sigma_{\rm v}^{xx}=0$.  This is the
superconducting (vortex glass) phase with $R(T=0)=0$.  With increasing
field and density, the vortices can delocalize at a field denoted by
$B_2$ in Fig.~\ref{fig1}, and will quantum diffuse with nonzero
diagonal $\sigma_{\rm v}^{xx}$ and nonzero electrical resistance - this is
the ``vortex metal.''  Finally, at large fields ($B_3$) the bosonic
vortices can condense giving $\sigma_{\rm v}^{xx} =  \infty$ (the
fermionized vortices form the $\nu=1$ IQHE state with
$\sigma^{xx}_{{\rm fv}} =0$).  This is the conventional Fermi liquid with $R
= 1/\sigma_{\rm s}$.  The spinon conductivity is likewise expected to be
zero at very low magnetic fields, since the spinons will predominantly
be localized at the cores of the well separated vortices.  At some
field, $B_1$, spinons may become delocalized, and form an impurity
band connecting vortices~\cite{smitha}. If this occurs with
$\sigma_{\rm v}^{xx} =0$, this is a ``superconducting thermal metal" phase.
Finally, in the very high field limit, one expects the spinon
resistivity to approach the normal Fermi liquid resistance,
$(\sigma_{\rm s}^{xx})^{-1} = R_N$.

For very disordered films with weakened superconductivity, the
vortices will be especially mobile, and will condense at low magnetic
fields,  $B_3 \ll H_{\rm c2}$.  With dilute vortices the spinon resistivity
could be significantly larger than $R_N$ at $B \approx B_3$, and the
films electrical resistance will have a very large peak just below
$B_3$.  This behavior is consistent to that observed in such
disordered samples.




Let us now consider the thermal conductivity $\kappa^{xx}$. We
assume that Fermi-liquid theory applies separately to both the
fermionized  vortices and spinons, then $\kappa^{xx} =
T(\sigma_{{\rm fv}}^{xx} + \sigma_{\rm s}^{xx})$ (setting the Lorentz ratio to be
1 for each fluid). Thus the Lorentz ratio for the two-fluid
vortex metal is $L=\kappa^{xx}/\sigma^{xx} T= (\sigma_{\rm s}^{xx} +
\sigma_{\rm fv}^{xx})/ [ \sigma_{\rm s}^{xx} + (\sigma_{\rm v}^{xx})^{-1} ]$.  This
violation of the Wiedemann-Franz law is a striking signature of the
non-Fermi liquid nature of the vortex metal.  Notice that $L$ varies
from zero to one as the magnetic field is tuned through the ``vortex
metal'' phase from $B_2$ up to $B_3$. Thus, the ``vortex metal''
naturally interpolates between the superconducting and normal 
phases.  As for the Nernst effect, ${\cal N} = {\cal E}_y/\partial_x
T$, with $\vec{J}=0$, ignoring small contributions from the
off-diagonal conductivities we find, ${\cal N} =
(\lambda_{\rm v}^{xx}/\sigma_{\rm v}^{xx}) R$. Within Fermi liquid theory (applied
to the  fermionic vortex fluid) $\lambda_{\rm v}^{xx} \propto T
\sigma_{{\rm fv}}^{xx}$, the Nernst signal is predicted to exhibit a
non-monotonic $B-$field dependence.


To address the Hall response, we need to consider non-zero
off-diagonal ``conductivities".  The Hall angle is then,
\begin{equation}
\label{Hall}
\tan{\theta_{\rm H}} = \frac{\sigma_{\rm v}^{xy} + A \sigma_{\rm
s}^{xy} } {\sigma_{\rm v}^{xx} + A \sigma_{\rm s}^{xx} };  \hskip1cm A
= (\sigma_{\rm v}^{xx})^2 + (\sigma_{\rm v}^{xy} )^2 .
\vspace*{-0.05in}
\end{equation}
Another quantity of interest is the Nernst angle.
Again ignoring small off-diagonal contributions, we get
\begin{equation}
\label{Nernst}
\tan {\theta_{\rm N}} = \frac{\lambda_{\rm s}^{xx}
\sigma_{\rm v}^{xx}}{\lambda_{\rm v}^{xx}} .
\vspace*{-0.05in}
\end{equation}
Thus the transport angles $\theta_{\rm H},\,\theta_{\rm N}$ differ in
the ``vortex metal".


Next we consider the specific heat, $C(T)$, in the ``vortex metal" at
low temperatures.  Naively, the vortex and spinon Fermi seas would
give the usual $C \sim T$ behavior.  But this is modified due to gauge
fluctuations.  Within RPA we integrate out the (diffusive) fermionic
spinons and vortices to obtain the quadratic action, $S_{\rm eff}= S_a +
S_{\rm int} + S_{\rm RPA}$ with, $S_{\rm RPA} = \int_{{\bf q},\omega_n} [ \rho_{\rm v} |
a_0-\alpha_0|^2 + {\cal D}_{\rm v} |a_t - \alpha_t|^2 + \rho_{\rm s} |\beta_0|^2 +
{\cal D}_{\rm s} |\beta_t|^2  ]$, where $a_t,\alpha_t$, and $\beta_t$ denote
transverse gauge fields in the Coulomb gauge, $\rho_{\rm v/s}$ are the
vortex/spinon density of states and ${\cal D}_{\rm v/s}({\bf q},\omega_n)
= \sigma_{\rm  v/s} | \omega_n| + \chi_{\rm v/s} q^2$, with ``magnetic
susceptibilities" $\chi_{\rm v/s}$.  The free energy can now be readily
extracted from the effective quadratic action giving, $ C(T) \sim T \ln\left( 1/T \right)  + {\cal O}(T)$.

Let us now discuss tunneling properties. Electron tunneling from a metallic tip
is expected to be similar to that in a conventional $2d$ metal due to the gapless
electron-like quasiparticles present in the vortex metal state. 
With a superconducting tip placed above the film, one can measure the  Cooper pair
tunneling conductance into the vortex metal~\cite{Naaman}.
The conductance $G = G_{\rm  A}+G_{\rm cp}$ has  two parts:  $G_{\rm A}$ is due to
simultaneous tunneling of two electrons, which is equivalent to the
Andreev reflection of an incident hole,  and $G_{\rm cp}$ is due to
tunneling of Cooper-pairs that get absorbed in the Cooper-pair
fluid. Whereas $G_{\rm A}$ measures the local density of states  of
the gapless quasiparticles and is almost temperature
independent,  $G_{\rm cp}$ depends strongly on temperature. $G_{\rm cp}$ is given by 
the Kubo formula, $G_{\rm cp}  \propto i \int dt t \left\langle \left[ I(t),\,
I(0) \right] \right\rangle$.  With the tip at  ${\bf x}={\bf 0}$, the
Cooper-pair current operator is $I= 2 e J \sin \phi_{\bf 0}$.  First,
we evaluate the imaginary time correlator,
$C(\tau) = \left\langle e^{i \phi_{\bf 0}(\tau)} e^{-i \phi_{\bf
0}(0)} \right\rangle$.
The density of Cooper pairs is the ``dual magnetic field,'' $b =
\epsilon_{ij} \partial_i a_j$ and the Cooper pair current is
${\hat{z}} \times {\bf e}$ with ``electric field'', $ {\bf e} = -
{\bm \nabla} a_0 - \partial_0 {\bf a}$.  $C(\tau)$ describes the
process of inserting a monopole at ${\bf x}={\bf 0}$ and removing it at a
later time $\tau$. The action of this monopole configuration
determines $C(\tau)$; we calculate it within a quadratic action
for the gauge field. Possibly, higher order terms in the gauge field
action affect the result, but they are beyond the scope of this
paper. Since the spinons do not affect the dominant large $\tau$
behavior~\cite{GalitskiRefaelFisher2}, we focus on the action
due to vortices.  Integrating them out within RPA, we obtain
\begin{equation}
\label{2+1}
S_0 = \frac{T}{2} \sum_{\omega_n} \int \frac{ d^2{\bf q}}{\left( 2 \pi
\right)^2} \left[ \varepsilon_{\alpha \beta} e_\alpha e_\beta + \mu
b^2 \right] + S_{\lambda}.
\vspace*{-0.05in}
\end{equation}
$S_0$ is expressed using the gauge invariant fields ${\bf e}$ and $b$.
Here $S_{\lambda} = i \int dx \lambda(x) \left[ {\bm \nabla} \times
{\bf e} + \partial_0 b - \rho_{\rm m}(x) \right]$ enforces the dual Maxwell
equation, with $\lambda(x)$ a Lagrange multiplier and $\rho_{\rm m}(x)$ the
density of monopoles;  we set $\rho_{\rm m}(x_0,{\bf r}) = \delta(x_0)
\delta({\bf r}) - \delta(x_0 - \tau) \delta({\bf r})$.  In
Eq.~(\ref{2+1}), $\varepsilon_{\alpha \beta}$ is the dielectric
constant determined by the diffusive fermionic vortices, with 
transverse part $\varepsilon^{({\rm tr})} = \sigma_{\rm v}/|\omega_n|$.
The permeability $\mu = \chi_{\rm v}$ is the vortex orbital
``magnetic'' susceptibility.  Using Eq.~(\ref{2+1}) we calculate the
action of the monopole configuration and find, $S(\tau) =
\frac{\sigma_{\rm v}}{2} \ln^2 \left( {\tau}/{\tau_0} \right)$, with
$\tau_0^{-1}$ the scattering rate of the diffusive vortices. 
Inserting the correlator $C(\tau) \propto \exp\left[-S(\tau)\right]$
into the Kubo formula and performing analytical continuation we find:
\begin{equation}
\label{tc}
G_{\rm cp}(T) \propto T^{-2} \exp{ \left[- \frac{\sigma_{\rm v}}{2} \ln^2
\left( T {\tau_0} \right) \right]} .
\vspace*{-0.05in}
\end{equation}
Note that the Andreev part of the conductance, $G_{\rm A}$, which describes two electron tunneling,
adds a $T$-independent contribution to $G$, which traces $\sigma_{\rm s}$:
\hspace*{3mm}$G_{\rm A}\propto \sigma_{\rm s}^2$.\hspace*{3mm} 
Near the superconducting transition, $B_2$,  $G_{\rm cp}$ is expected to dominate the
tunneling since $\sigma_{\rm v}$ is small. Particularly, when vortices are localized 
$G_{\rm cp}$ diverges, which corresponds to the appearance of the Josephson effect.
As the magnetic field, $\sigma_{\rm v}$, and $\sigma_{\rm s}$ grow,
$G_{\rm A}$ will increase and the temperature dependence of $G$ gets
suppressed. Therefore, tunneling provides another probe of the
superconductor-insulator cross-over in the ``vortex metal.''  

We can finally address the effects of ${\cal L}_\Delta$ in
Eq.~(\ref{Ldel}) in the ``vortex metal'' within the RPA. The
perturbative effects of ${\cal L}_\Delta$ can be extracted from the
two-point correlation function of $f_\uparrow f_\downarrow {\cal
O}_\alpha$ evaluated at $\Delta=0$.  Equivalently we can consider the correlators of $e^{i\phi} c_{\uparrow}c_{\downarrow}$. 
Assuming factorization to an electron correlator times a Cooper pair correlator, the electrons contribute a power law since $\langle c_\alpha (\tau) c^\dagger_\alpha (0)\rangle \sim \tau^{-1}$.
The imaginary time correlator is
$C(\tau) \propto \exp\left[-\sigma_{\rm v} \ln^2(\tau/\tau_0) \right]$. Since this decays faster than
any power, we expect that in the ``vortex metal'' a small $\Delta$
is perturbatively irrelevant in the RG sense.

In this Letter we introduced a new two fluid description for
amorphous superconducting films in a magnetic field.
Fermionization of the field-induced vortices allowed us to access a novel
resistive quantum-liquid phase that naturally interpolates between
the low $B$ superconductor and the normal metal for $B \ge H_{\rm c2}$.
Anomalous thermoelectric transport and tunneling behavior were predicted in the
 ``vortex metal."  It is our hope that this paper will help
 motivate new experiments on such 2d amorphous films.

It is a pleasure to thank A.~Auerbach, L.~Balents, A. Kapitulnik, S.~Kivelson, A.~I.~Larkin, O.~Motrunich,
Y.~Oreg, and D.~Shahar for discussions on this work. This work was generously supported by the  David and Lucile
Packard foundation (V.G.) and by the National Science Foundation through
grants  PHY99-07949 (V.G., G.R., and M.P.A.F.), DMR-0210790 (M.P.A.F.), and DMR-0308945 (T.S.).
T.S. also acknowledges
funding from the NEC Corporation, the Alfred P. Sloan Foundation, and an award from The Research Corporation.
\vspace*{-0.2in}

\bibliography{VFL}

\begin{thebibliography}{20}
\expandafter\ifx\csname natexlab\endcsname\relax\def\natexlab#1{#1}\fi
\expandafter\ifx\csname bibnamefont\endcsname\relax
  \def\bibnamefont#1{#1}\fi
\expandafter\ifx\csname bibfnamefont\endcsname\relax
  \def\bibfnamefont#1{#1}\fi
\expandafter\ifx\csname citenamefont\endcsname\relax
  \def\citenamefont#1{#1}\fi
\expandafter\ifx\csname url\endcsname\relax
  \def\url#1{\texttt{#1}}\fi
\expandafter\ifx\csname urlprefix\endcsname\relax\def\urlprefix{URL }\fi
\providecommand{\bibinfo}[2]{#2}
\providecommand{\eprint}[2][]{\url{#2}}

\bibitem[{\citenamefont{Haviland et~al.}(1989)\citenamefont{Haviland, Liu, and
  Goldman}}]{Goldman1}
\bibinfo{author}{\bibfnamefont{D.~B.} \bibnamefont{Haviland}},
  \bibinfo{author}{\bibfnamefont{Y.}~\bibnamefont{Liu}}, \bibnamefont{and}
  \bibinfo{author}{\bibfnamefont{A.~M.} \bibnamefont{Goldman}},
  \bibinfo{journal}{Phys. Rev. Lett.} \textbf{\bibinfo{volume}{62}},
  \bibinfo{pages}{2180} (\bibinfo{year}{1989}).

\bibitem[{\citenamefont{Hebard and Paalanen}(1990)}]{Hebard}
\bibinfo{author}{\bibfnamefont{A.~F.} \bibnamefont{Hebard}} \bibnamefont{and}
  \bibinfo{author}{\bibfnamefont{M.~A.} \bibnamefont{Paalanen}},
  \bibinfo{journal}{Phys. Rev. Lett.} \textbf{\bibinfo{volume}{65}},
  \bibinfo{pages}{927} (\bibinfo{year}{1990}).

\bibitem[{\citenamefont{Gantmakher and et~al.}(1998)}]{Gantmakher}
\bibinfo{author}{\bibfnamefont{V.~F.} \bibnamefont{Gantmakher}}
  \bibnamefont{and} \bibinfo{author}{\bibnamefont{et~al.}},
  \bibinfo{journal}{JETP Lett.} \textbf{\bibinfo{volume}{68}},
  \bibinfo{pages}{363} (\bibinfo{year}{1998}).

\bibitem[{\citenamefont{Fisher}(1990)}]{Fisher}
\bibinfo{author}{\bibfnamefont{M.~P.~A.} \bibnamefont{Fisher}},
  \bibinfo{journal}{Phys. Rev. Lett.} \textbf{\bibinfo{volume}{65}},
  \bibinfo{pages}{923} (\bibinfo{year}{1990}).

\bibitem[{\citenamefont{Mason and Kapitulnik}(1999)}]{Kapitulnik1}
\bibinfo{author}{\bibfnamefont{N.}~\bibnamefont{Mason}} \bibnamefont{and}
  \bibinfo{author}{\bibfnamefont{A.}~\bibnamefont{Kapitulnik}},
  \bibinfo{journal}{Phys. Rev. Lett.} \textbf{\bibinfo{volume}{82}},
  \bibinfo{pages}{5341} (\bibinfo{year}{1999}).

\bibitem[{\citenamefont{Mason and Kapitulnik}(2001)}]{Kapitulnik2}
\bibinfo{author}{\bibfnamefont{N.}~\bibnamefont{Mason}} \bibnamefont{and}
  \bibinfo{author}{\bibfnamefont{A.}~\bibnamefont{Kapitulnik}},
  \bibinfo{journal}{Phys. Rev. B} \textbf{\bibinfo{volume}{64}},
  \bibinfo{pages}{060504} (\bibinfo{year}{2001}).

\bibitem[{\citenamefont{Palaanen et~al.}(1992)\citenamefont{Palaanen, Hebard,
  and Ruel}}]{HPR}
\bibinfo{author}{\bibfnamefont{M.~A.} \bibnamefont{Palaanen}},
  \bibinfo{author}{\bibfnamefont{A.~F.} \bibnamefont{Hebard}},
  \bibnamefont{and} \bibinfo{author}{\bibfnamefont{R.~R.} \bibnamefont{Ruel}},
  \bibinfo{journal}{Phys. Rev. Lett.} \textbf{\bibinfo{volume}{69}},
  \bibinfo{pages}{1609} (\bibinfo{year}{1992}).

\bibitem[{\citenamefont{Sambandamurthy
  et~al.}(2004)\citenamefont{Sambandamurthy, Engel, Johansson, and
  Shahar}}]{Shahar}
\bibinfo{author}{\bibfnamefont{G.}~\bibnamefont{Sambandamurthy}},
  \bibinfo{author}{\bibfnamefont{L.~W.} \bibnamefont{Engel}},
  \bibinfo{author}{\bibfnamefont{A.}~\bibnamefont{Johansson}},
  \bibnamefont{and} \bibinfo{author}{\bibfnamefont{D.}~\bibnamefont{Shahar}},
  \bibinfo{journal}{Phys. Rev. Lett.} \textbf{\bibinfo{volume}{92}},
  \bibinfo{pages}{107005} (\bibinfo{year}{2004}).

\bibitem[{\citenamefont{Steiner et~al.}(2005)\citenamefont{Steiner, Boebinger,
  and Kapitulnik}}]{KapitulnikSteiner}
\bibinfo{author}{\bibfnamefont{M.~A.} \bibnamefont{Steiner}},
  \bibinfo{author}{\bibfnamefont{G.}~\bibnamefont{Boebinger}},
  \bibnamefont{and}
  \bibinfo{author}{\bibfnamefont{A.}~\bibnamefont{Kapitulnik}},
  \bibinfo{journal}{Phys. Rev. Lett.} \textbf{\bibinfo{volume}{94}},
  \bibinfo{pages}{107008} (\bibinfo{year}{2005}).

\bibitem[{\citenamefont{Dalidovic and Phillips}(2001)}]{Phillips}
\bibinfo{author}{\bibfnamefont{D.}~\bibnamefont{Dalidovic}} \bibnamefont{and}
  \bibinfo{author}{\bibfnamefont{P.}~\bibnamefont{Phillips}},
  \bibinfo{journal}{Phys. Rev. B} \textbf{\bibinfo{volume}{64}},
  \bibinfo{pages}{052507} (\bibinfo{year}{2001}).

\bibitem[{\citenamefont{Dalidovic and Phillips}(2002)}]{PhillipsPRL}
\bibinfo{author}{\bibfnamefont{D.}~\bibnamefont{Dalidovic}} \bibnamefont{and}
  \bibinfo{author}{\bibfnamefont{P.}~\bibnamefont{Phillips}},
  \bibinfo{journal}{Phys. Rev. Lett.} \textbf{\bibinfo{volume}{89}},
  \bibinfo{pages}{027001} (\bibinfo{year}{2002}).

\bibitem[{\citenamefont{Kapitulnik et~al.}(2001)\citenamefont{Kapitulnik,
  Mason, Kivelson, and Chakravarty}}]{Kivelson}
\bibinfo{author}{\bibfnamefont{A.}~\bibnamefont{Kapitulnik}},
  \bibinfo{author}{\bibfnamefont{N.}~\bibnamefont{Mason}},
  \bibinfo{author}{\bibfnamefont{S.~A.} \bibnamefont{Kivelson}},
  \bibnamefont{and}
  \bibinfo{author}{\bibfnamefont{S.}~\bibnamefont{Chakravarty}},
  \bibinfo{journal}{Phys. Rev. B} \textbf{\bibinfo{volume}{63}},
  \bibinfo{pages}{125322} (\bibinfo{year}{2001}).

\bibitem[{\citenamefont{Feigelman et~al.}(1993)\citenamefont{Feigelman,
  Geshkenbein, Ioffe, and Larkin}}]{Larkin}
\bibinfo{author}{\bibfnamefont{M.~V.} \bibnamefont{Feigelman}},
  \bibinfo{author}{\bibfnamefont{V.~B.} \bibnamefont{Geshkenbein}},
  \bibinfo{author}{\bibfnamefont{L.~B.} \bibnamefont{Ioffe}}, \bibnamefont{and}
  \bibinfo{author}{\bibfnamefont{A.~I.} \bibnamefont{Larkin}},
  \bibinfo{journal}{Phys. Rev. B} \textbf{\bibinfo{volume}{48}},
  \bibinfo{pages}{16641} (\bibinfo{year}{1993}).

\bibitem[{\citenamefont{Senthil and Fisher}(2000)}]{SenthilFisher}
\bibinfo{author}{\bibfnamefont{T.}~\bibnamefont{Senthil}} \bibnamefont{and}
  \bibinfo{author}{\bibfnamefont{M.~P.~A.} \bibnamefont{Fisher}},
  \bibinfo{journal}{Phys. Rev. B} \textbf{\bibinfo{volume}{62}},
  \bibinfo{pages}{7850} (\bibinfo{year}{2000}).

\bibitem[{\citenamefont{Balents and Fisher}(2005)}]{BalentsFisher}
\bibinfo{author}{\bibfnamefont{L.}~\bibnamefont{Balents}} \bibnamefont{and}
  \bibinfo{author}{\bibfnamefont{M.~P.~A.} \bibnamefont{Fisher}},
  \bibinfo{journal}{Phys. Rev. B} \textbf{\bibinfo{volume}{71}},
  \bibinfo{pages}{085119} (\bibinfo{year}{2005}).

\bibitem[{\citenamefont{Read}(1998)}]{Read}
\bibinfo{author}{\bibfnamefont{N.}~\bibnamefont{Read}}, \bibinfo{journal}{Phys.
  Rev. B} \textbf{\bibinfo{volume}{58}}, \bibinfo{pages}{16262}
  (\bibinfo{year}{1998}).

\bibitem[{\citenamefont{Alicea et~al.}()\citenamefont{Alicea, Motrunich,
  Hermele, and Fisher}}]{AliceaMotrunich}
\bibinfo{author}{\bibfnamefont{J.}~\bibnamefont{Alicea}},
  \bibinfo{author}{\bibfnamefont{O.~I.} \bibnamefont{Motrunich}},
  \bibinfo{author}{\bibfnamefont{M.}~\bibnamefont{Hermele}}, \bibnamefont{and}
  \bibinfo{author}{\bibfnamefont{M.~P.~A.} \bibnamefont{Fisher}},
  \bibinfo{howpublished}{cond-mat/0503399}.

\bibitem[{\citenamefont{Vishveshwara et~al.}(2000)\citenamefont{Vishveshwara,
  Senthil, and Fisher}}]{smitha}
\bibinfo{author}{\bibfnamefont{S.}~\bibnamefont{Vishveshwara}},
  \bibinfo{author}{\bibfnamefont{T.}~\bibnamefont{Senthil}}, \bibnamefont{and}
  \bibinfo{author}{\bibfnamefont{M.~P.~A.} \bibnamefont{Fisher}},
  \bibinfo{journal}{PRB} \textbf{\bibinfo{volume}{61}}, \bibinfo{pages}{6966}
  (\bibinfo{year}{2000}).

\bibitem[{\citenamefont{Naaman et~al.}(2001)\citenamefont{Naaman, Teizer, and
  Dynes}}]{Naaman}
\bibinfo{author}{\bibfnamefont{O.}~\bibnamefont{Naaman}},
  \bibinfo{author}{\bibfnamefont{W.}~\bibnamefont{Teizer}}, \bibnamefont{and}
  \bibinfo{author}{\bibfnamefont{R.}~\bibnamefont{Dynes}},
  \bibinfo{journal}{Phys. Rev. Lett.} \textbf{\bibinfo{volume}{87}},
  \bibinfo{pages}{097004} (\bibinfo{year}{2001}).

\bibitem[{\citenamefont{Refael et~al.}()\citenamefont{Refael, Galitski, and
  Fisher}}]{GalitskiRefaelFisher2}
\bibinfo{author}{\bibfnamefont{G.}~\bibnamefont{Refael}},
  \bibinfo{author}{\bibfnamefont{V.~M.} \bibnamefont{Galitski}},
  \bibnamefont{and} \bibinfo{author}{\bibfnamefont{M.~P.~A.}
  \bibnamefont{Fisher}}, \bibinfo{howpublished}{in preparation}.

\end{thebibliography}

\end{document}